\documentclass[a4paper]{jpconf}
\usepackage{graphicx}
\usepackage{slashbox}
\begin{document}
\title{Failure of Breit-Wigner and success of dispersive descriptions of the $\tau^-\to K^-\eta\nu_\tau$ decays}

\author{Pablo Roig}

\address{Instituto de F\'{\i}sica, Universidad Nacional Aut\'onoma de M\'exico, Apartado Postal 20-364, 01000 M\'exico D.F., M\'exico}

\ead{pabloroig@fisica.unam.mx}

\begin{abstract}
The $\tau^-\to K^-\eta\nu_\tau$ decays have been studied using Chiral Perturbation Theory extended by including resonances as active fields. We have found that the 
treatment of final state interactions is crucial to provide a good description of the data. The Breit-Wigner approximation does not resum them and neglects the real 
part of the corresponding chiral loop functions, which violates analyticity and leads to a failure in the confrontation with the data. On the contrary, its resummation by 
means of an Omn\`es-like exponentiation of through a dispersive representation provides a successful explanation of the measurements. These results illustrate the fact that 
Breit-Wigner parametrizations of hadronic data, although simple and easy to handle, lack a link with the underlying strong interaction theory and should be avoided.\\

As a result of our analysis we determine the properties of the $K^\star(1410)$ resonance with a precision competitive to its traditional extraction using 
$\tau^-\to (K\pi)^-\nu_\tau$ decays, albeit the much limited statistics accumulated for the $\tau^-\to K^-\eta\nu_\tau$ channel. We also predict the soon discovery of the 
$\tau^-\to K^-\eta^\prime\nu_\tau$ decays.
\end{abstract}

\section{Introduction}
\vspace*{0.25cm}
Hadronic tau decays provide a clean laboratory to test the non-perturbative strong interaction involving mesons in rather clean conditions \cite{HadTauDec}. At the inclusive 
level this allows to extract fundamental parameters of the Standard Theory, like the strong coupling at the tau mass scale, the CKM matrix element $V_{us}$ or the mass of 
the strange quark \cite{InclTauDec}. On the exclusive side, the non-trivial strangeness-changing processes studied more accurately are the $\tau^-\to(K\pi)^-\nu_\tau$ 
decays, even more with the advent of the B-factory measurements performed by the BaBar and Belle Collaborations \cite{BfactKpi}. These very precise data have triggered a number 
of dedicated theoretical studies \cite{Kpi} which have taken advantage of the $K^\star(892)$-exchange dominance in these decays to determine very precisely the mass, width 
and couplings of this resonance. Although subleading, the contribution of its first excitation, the $K^\star(1410)$ meson, can also be probed through the 
$K\pi$ tau decays. As a result, its parameters were also evaluated in the quoted studies, even though with much less precision than in the case of its lightest copy.\\

Phase space considerations suggest that the related $\tau^-\to K^-\eta\nu_\tau$ decays may be specially sensitive to the properties of the $K^\star(1410)$ resonance and to 
its interplay with the $K^\star(892)$ meson. To test this possibility was one of the motivations of our study \cite{OurWork} aiming to evaluate the hadronic matrix element 
and give sensible estimates of invariant mass spectrum and decay width that could be confronted to the BaBar and Belle measurements \cite{BfactKeta}. We also wanted to 
improve the description of the strange spectral function with a better understanding of this decay channel and to provide TAUOLA \cite{TAUOLA}, the standard Monte Carlo 
generator for tau lepton decays, with theory-based hadronic currents \cite{Olga} for this process that, up to know, was modeled very crudely using a constant form factor. In 
this sense, Belle's paper requested more elaborated theoretical analyses of tau decays including $\eta$ mesons beyond the classic works \cite{TauEta} to match the precision 
of current measurements. This improvement is also requested from the Monte Carlo point of view \cite{Actis:2010gg}.\\

BaBar and Belle measurements \cite{BfactKeta} of the $\tau^-\to K^-\eta\nu_\tau$ decays yield the respective branching fractions $(1.58\pm0.05\pm0.09)\cdot10^{-4}$ and 
$(1.42\pm0.11\pm0.07)\cdot10^{-4}$. The partner $\tau^-\to K^- \eta^\prime \nu_\tau$ decays have not been detected yet, but an upper branching ratio of $2.4\cdot10^{-6}$ at 
the $90\%$ confidence level was placed by BaBar \cite{Lees:2012ks}. Branching ratios at the level of $10^{-6}$ should be within reach at the forthcoming Belle-II experiment. 
Therefore, another target of our analysis was to determine if a soon discovery of this decay mode could be expected.\\

\section{Matrix element and decay width}
\vspace*{0.25cm}
The differential decay width for the considered processes reads
\begin{eqnarray} \label{spectral function}
& & \frac{d\Gamma\left(\tau^-\to K^-\eta^{(\prime)}\nu_\tau\right)}{d\sqrt{s}} = \frac{G_F^2M_\tau^3}{32\pi^3s}S_{EW}\Big|V_{us}f_+^{K^-\eta^{(\prime)}}(0)\Big|^2
\left(1-\frac{s}{M_\tau^2}\right)^2\\
& & \left\lbrace\left(1+\frac{2s}{M_\tau^2}\right)q_{K\eta^{(\prime)}}^3(s)\Big|\widetilde{f}_+^{K^-\eta^{(\prime)}}(s)\Big|^2+\frac{3\Delta_{K\eta^{(\prime)}}^2}{4s}q_{K\eta^{(\prime)}}(s)\Big|\widetilde{f}_0^{K^-\eta^{(\prime)}}(s)\Big|^2\right\rbrace\,,\nonumber
\end{eqnarray}
where
\begin{eqnarray}\label{definitions}
& &  s=\left(p_{\eta^{(\prime)}}+p_K\right)^2 \,,\quad q_{PQ}(s)=\frac{\sqrt{s^2-2s\Sigma_{PQ}+\Delta_{PQ}^2}}{2\sqrt{s}}\,,
\nonumber\\
& & \Sigma_{PQ}=m_P^2+m_Q^2\,,\quad \Delta_{PQ}=m_P^2-m_Q^2\,
,\quad \widetilde{f}_{+,0}^{K^-\eta^{(\prime)}}(s)=\frac{f_{+,0}^{K^-\eta^{(\prime)}}(s)}{f_{+,0}^{K^-\eta^{(\prime)}}(0)}\,,
\end{eqnarray}
and $S_{EW} = 1.0201$ \cite{Erler:2002mv} represents an electroweak correction factor. $\Big|V_{us}f_+^{K^-\eta}(0)\Big|=\Big|V_{us}f_+^{K^-\pi^0}(0)\mathrm{cos}\theta_P\Big|$ 
and $\Big|V_{us}f_+^{K^-\eta^\prime}(0)\Big|=\Big|V_{us}f_+^{K^-\pi^0}(0)\mathrm{sin}\theta_P\Big|$, where $\theta_P=(-13.3\pm1.0)^\circ$ \cite{Ambrosino:2006gk}. From these 
values we can already guess the suppression of the $\tau^-\to K^-\eta^\prime\nu_\tau$ decays. Unitarization effects will increase the scalar form factor contribution to this 
decay channel making it more important than the vector form factor effect, but it will still be roughly two orders of magnitude less frequent than the $\tau^-\to K^-\eta\nu_\tau$ 
decays. We will use the value $\Big|V_{us}f_+^{K^-\pi^0}(0)\Big|=0.21664\pm0.00048$, obtained analyzing semi-leptonic Kaon decay data~\cite{Antonelli:2010yf}.\\

In eq.~(\ref{spectral function}) the strong interaction dynamics has been encoded in the $f_{+,0}^{K^-\eta^{(\prime)}}(0)$ form factors defined by
\begin{equation}\label{Had m.e.}
 \left\langle K^-\eta^{(\prime)} \Big|\bar{s}\gamma^\mu u\Big| 0\right\rangle=\left[\left(p_{\eta^{(\prime)}}-p_K\right)^\mu +\frac{\Delta_{K \eta^{(\prime)}}}{s}q^\mu\right]c^V_{K^-\eta^{(\prime)}}f_+^{K^-\eta^{(\prime)}}(s)+
\frac{\Delta_{K \pi}}{s}q^\mu c^S_{K^-\eta^{(\prime)}} f_0^{K^-\eta^{(\prime)}}(s)\,,
\end{equation}
with the normalization condition
\begin{equation}\label{condition origin}
 f_+^{K^-\eta^{(\prime)}}(0)=-\frac{c^S_{K^-\eta^{(\prime)}}}{c^V_{K^-\eta^{(\prime)}}}\frac{\Delta_{K\pi}}{\Delta_{K\eta^{(\prime)}}}f_0^{K^-\eta^{(\prime)}}(0)\,
\end{equation}
and the values $c^V_{K\eta^{(\prime)}}=-\sqrt{\frac{3}{2}}$, $c^S_{K^-\eta}= -\frac{1}{\sqrt{6}}$, $c^S_{K^-\eta^\prime}= \frac{2}{\sqrt{3}}$.
The use of the tilded $\widetilde{f}_{+,0}^{K^-\eta^{(\prime)}}(s)$ form factors in eq.~(\ref{spectral function}) yields more compact expressions than previously used, which 
are explicitly symmetric under the exchange $\eta\leftrightarrow\eta^\prime$.\\

\section{Scalar and vector form factors in Resonance Chiral Theory}
\vspace*{0.25cm}
Nowadays, there is not any analytic way known of obtaining the relevant $\widetilde{f}_{+,0}^{K^-\eta^{(\prime)}}(s)$ form factors employing the quantum field theory of 
strong interactions, QCD~\footnote{Lattice simulations do not provide these form factors at the moment.}. This, however, does not mean that its Lagrangian is useless for 
obtaining them. In particular, at very low energies, QCD exhibits and approximate (chiral) symmetry in the limit of massless light quarks. This property allows to build 
an effective field theory dual to it in this regime, Chiral Perturbation Theory ($\chi PT$) \cite{ChPT}.\\

However, as the energy increases, it does not suffice to include higher and higher order computations in $\chi PT$ \cite{ChPT2}. On the contrary, it is necessary to incorporate 
the next lighter states, the lowest-lying light-flavoured resonances, as active degrees of freedom into the action. In the context of tau decays this feature was studied in 
Ref.~\cite{Colangelo:1996hs}.\\

A complementary and equivalent view comes from considering the expansion parameter of $\chi PT$, the ratio between momenta and masses of the lightest pseudoscalar ($\pi$, $K$ 
and $\eta$) mesons over the chiral symmetry breaking scale, of order $1$ GeV. It is obvious that when its value starts to be comparable to one the chiral expansion will no 
longer converge. This scale is (parametrically) of the same order of the mass of the lightest mesons, i.e. the $\rho(770)$ resonance. Therefore, the need of finding 
an alternative expansion parameter valid in the GeV-region also leads to an extension of $\chi PT$. A successful candidate was provided by the $1/N_C$ expansion of QCD \cite{LargeNc}, 
whose predictions for a theory including resonances are corroborated experimentally \cite{LargeNc2}.\\

A realization of these ideas is provided by the Resonance Chiral Theory ($R\chi T$) \cite{RChT}, which includes the meson resonances in the antisymmetric tensor field representation. 
This formalism brings in the advantage that there is no need to include the local contact interactions at subleading orders in $\chi PT$ because they are reproduced upon 
integrating the resonances out. The values of the $R\chi T$ couplings are not restricted by symmetry requirements. However, the resonance Physics is supposed to interpolate 
between the known chiral and perturbative regimes. Consequently, the matching between the operator product expansion and the $R\chi T$ results for Green functions is 
performed \cite{RChT, GFs}, rendering relations among and predictions of the Lagrangian couplings which increase the predictability of the theory. Remarkably, within $R\chi T$ 
the successful notion of vector meson dominance is not an \textit{a priori} assumption but a dynamical result. The application of $R\chi T$ to the study of hadronic tau 
decays has proved successful in a variety of decay channels \cite{Kpi, OurWork, Olga, HadTauDec2} and related processes \cite{FFs}.\\

The relevant effective Lagrangian for the lightest resonance nonets is
\begin{eqnarray}
\label{eq:ret} {\cal L}_{\rm R\chi T}   & \doteq   & {\cal L}_{\rm kin}^{\rm V,S}\, + \, \frac{F^2}{4}\langle u_{\mu} u^{\mu} + \chi _+\rangle \, + \, 
\frac{F_V}{2\sqrt{2}} \langle V_{\mu\nu} f_+^{\mu\nu}\rangle\,+\,i \,\frac{G_V}{\sqrt{2}} \langle V_{\mu\nu} u^\mu u^\nu\rangle \,+\, c_d \langle S u_{\mu} u^{\mu}\rangle 
\,+\, c_m \langle S \chi_ +\rangle\,,\quad\quad
\label{lagrangian}
\end{eqnarray}
where all coupling constants are real: $F$ is the pion decay constant, $F_V$ ($c_m$) gives the coupling of the vector (scalar) resonances to the $W$ current (scalar source) 
and $G_V$ ($c_d$) provides the coupling of the vector (scalar) mesons to pairs of pseudoscalars. The definition of the chiral tensors entering eq.~(\ref{lagrangian}) can be 
found in Refs.~\cite{RChT}.\\

The resulting vector form factors are
\begin{equation} \label{RChT VFFs}
 \tilde{f}_+^{K^-\eta}(s)=\frac{f_+^{K^-\eta}(s)}{f_+^{K^-\eta}(0)}=1+\frac{F_V G_V}{F^2}\frac{s}{M_{K^\star}^2-s}\,=\frac{f_+^{K^-\eta^\prime}(s)}{f_+^{K^-\eta^\prime}(0)}=\, \tilde{f}_+^{K^-\eta^\prime}(s)\,,
\end{equation}
where the values of $f_+^{K^-\eta^{(\prime)}}(0)$ were used.\\

The strangeness changing scalar form factors and associated S-wave scattering within $R\chi T$ have been investigated in a series of papers by Jamin, Oller and Pich 
\cite{JOP}. Our expressions for the scalar form factors can be written in terms of the $f_0^{K^-\eta_8}(s)$, $f_0^{K^-\eta_1}(s)$ form factors given in Refs.~\cite{JOP}. It 
should be noted that using our conventions, the tilded scalar form factors become simply
\begin{equation}\label{RChT SFFs def}
  \tilde{f}_0^{K^-\eta}(s)=\frac{f_0^{K^-\eta}(s)}{f_0^{K^-\eta}(0)}=1+\frac{c_d c_m}{4 F^2}\frac{s}{M_S^2-s}\,=\frac{f_+^{K^-\eta^\prime}(s)}{f_0^{K^-\eta^\prime}(0)}=\, \tilde{f}_0^{K^-\eta^\prime}(s)\,,
\end{equation}
which are more compact than the corresponding results in Refs.~\cite{JOP} and display the same symmetry $\eta\leftrightarrow\eta^\prime$ than the vector form factors in 
eq.~(\ref{RChT VFFs}).\\

The contribution of heavier resonances can be included within $R\chi T$ in the spirit of the $N_C\to\infty$ limit. However, within this infinite tower of states only the 
$K^\star(1410)$ resonance will play a relevant role in the considered processes (in the vector form factor). In the scalar one, it should be pointed out that the resonance 
labeled $S$ in eq.~(\ref{RChT SFFs def}) corresponds to the $K^\star_0(1430)$, since the $\kappa(800)$ meson is dynamically generated through $K\pi$ rescattering \cite{kappa}.\\
Then, the vector form factor in eq.~(\ref{RChT VFFs}) shall be replaced by
\begin{equation} \label{RChT VFFs2Res}
 \tilde{f}_+^{K^-\eta^{(\prime)}}(s)=1+\frac{F_V G_V}{F^2}\frac{s}{M_{K^\star}^2-s}+\frac{F_V^\prime G_V^\prime}{F^2}\frac{s}{M_{K^\star\prime}^2-s}\,,
\end{equation}
where the operators with couplings $F_V^\prime$ and $G_V^\prime$ are defined in analogy with the corresponding unprimed couplings in eq.~(\ref{lagrangian}).

The vanishing of the $f_+^{K^-\eta^{(\prime)}}(s)$ and $f_0^{K^-\eta^{(\prime)}}(s)$ form factors for $s\to\infty$ at least as $1/s$ \cite{BL} determines the short-distance 
constraints
\begin{equation}\label{shortdistance}
 F_V G_V + F_V^\prime G_V^\prime= F^2\,,\quad 4 c_d c_m = F^2\,,\quad c_d-c_m=0\,,
\end{equation}
which yield the form factors
\begin{eqnarray} \label{FFs with short distance constraints}
& & \tilde{f}_+^{K^-\eta}(s)=\frac{M_{K^\star}^2+\gamma s}{M_{K^\star}^2-s}-\frac{\gamma s}{M_{K^\star\prime}^2-s}= \tilde{f}_+^{K^-\eta^\prime}(s)\,,\\
& & \tilde{f}_0^{K^-\eta}(s) = \frac{M_S^2}{M_S^2-s}\,=\,\tilde{f}_0^{K^-\eta^\prime}(s)\,,\nonumber
\end{eqnarray}
where $\gamma=-\frac{F_V^\prime G_V^\prime}{F^2}=\frac{F_VG_V}{F^2}-1$ \cite{Kpi}. The modifications introduced by heavier resonances to this relation and to the definition 
of $\gamma$ are negligible.\\

\section{Treatment of final state interactions}
\vspace*{0.25cm}
The form factors in eq.~(\ref{FFs with short distance constraints}) diverge for $s=M^2$, while the hadronic observables are peaked (but not divergent) at these energies. The 
solution comes by including a subleading effect in the large-$N_C$ expansion as given by the finite width of the resonances. In fact, some of them are wide enough so that the 
energy dependence of their widths becomes an issue. This topic has been studied within $R\chi T$ in Ref.~\cite{GomezDumm:2000fz} were a precise and consistent definition for 
the spin-one meson off-shell widths was given. Its application to the $K^\star(892)$ case yields an expression that does quite a good job in the comparisons with data. However, 
at the present level of precision it is preferable to include the corrections to the corresponding on-shell width value to write
\begin{eqnarray}\label{K^* width}
 \Gamma_{K^*}(s) & = & \Gamma_{K^*}\left(M_{K^*}^2\right)\frac{s}{M_{K^*}^2}\frac{\sigma_{K\pi}^3(s)+\mathrm{cos}^2\theta_P \sigma_{K\eta}^3(s) + \mathrm{sin}^2\theta_P\sigma_{K\eta^\prime}^3(s)}{\sigma_{K\pi}^3(M_{K^*}^2)}\,,
\end{eqnarray}
where it was used that the only absorptive cut at the $M_{K^\star}$-scale is given by the elastic contribution and $\sigma_{PQ}(s)=\frac{2q_{PQ}(s)}{\sqrt{s}}\theta\left(s-(m_P+m_Q)^2\right)$.\\

Contrary to the $K^\star(892)$ case, there is no restriction from the chiral limit that applies to the width of the $K^\star(1410)$. Assuming that the lightest $K\pi$ absorptive 
contribution dominates one has
\begin{equation}\label{Kstarprimewidth}
 \Gamma_{K^{\star\prime}}(s)\,=\,\Gamma_{K^{\star\prime}}\left(M_{K^\star\prime}^2\right)\frac{s}{M_{K^{\star\prime}}^2}\frac{\sigma_{K\pi}^3(s)}{\sigma_{K\pi}^3(M_{K^{\star\prime}}^2)}\,.
\end{equation}
The scalar resonance width can also be computed in $R\chi T$ similarly yielding, for the $K^\star_0(1430)$
\begin{equation}\label{Gamma S computed}
 \Gamma_{S}(s)\,=\,\Gamma_{S}\left(M_S^2\right)\left(\frac{s}{M_S^2}\right)^{3/2}\frac{g(s)}{g\left(M_S^2\right)}\,,
\end{equation}
with
\begin{eqnarray}\label{g(s)}
 g(s) & = & \frac{3}{2}\sigma_{K\pi}(s)+\frac{1}{6}\sigma_{K\eta}(s)\left[\mathrm{cos}\theta_P\left(1+\frac{3\Delta_{K\pi}+\Delta_{K\eta}}{s}\right)+2\sqrt{2}\mathrm{sin}\theta_P\left(1+\frac{\Delta_{K\eta}}{s}\right)\right]^2\nonumber\\
& & +\frac{4}{3}\sigma_{K\eta^\prime}(s)\left[\mathrm{cos}\theta_P\left(1+\frac{\Delta_{K\eta^\prime}}{s}\right)-\frac{\mathrm{sin}\theta_P}{2\sqrt{2}}\left(1+\frac{3\Delta_{K\pi}+\Delta_{K\eta^\prime}}{s}\right)\right]^2\,.
\end{eqnarray}

The final state interactions (FSI) encoded in the chiral loop functions are not small and the real parts of these functions are not negligible. Moreover, even if they were 
numerically small, disregarding them while keeping the corresponding imaginary part (giving the meson widths) would violate analyticity. The key point is how to handle the 
resummation of these FSI. We will show that a Breit-Wigner approximation (which neglects the real part of the chiral loop functions and does not resum FSI) fails dramatically 
in the $\tau^-\to K^-\eta\nu_\tau$ decays. On the contrary, two kinds of resummations of FSI (Omn\`es-like and dispersive representation) do provide a good agreement with the 
data. The latter, which also gives the best results, is preferable because it is analytic by construction, while in the former analyticity only holds perturbatively. We 
emphasize that despite Breit-Wigner parametrizations of hadronic data are very simple and manageable, they violate known properties of the underlying strong interaction 
(analyticity, unitarity, asymptotic behaviour, ...). Therefore very little can be learned about QCD by using them. If that is our purpose (as it is supposed to be), they 
should not be employed.\\

Specifically, three different options will be considered, in increasing degree of soundness, for the treatment of FSI.  The relevant form factors will be obtained from 
eqs.(\ref{FFs with short distance constraints}) in each case by:
\begin{itemize}
 \item Dipole model (Breit-Wigner): $M_R^2-s$ will be replaced by $M_R^2-s-iM_R\Gamma_R(s)$ with $\Gamma_{K^\star}(s)$ and $\Gamma_S(s)$ given by eqs. (\ref{K^* width}) 
and (\ref{Gamma S computed}). Since the scalar form factors will not be unitarized, the $\kappa(800)$ resonance contribution is lost (or at best badly modeled) in this 
approach. Analyticity is violated already at the lowest order.
 \item Exponential parametrization (JPP): The Breit-Wigner vector form factor described above is multiplied by the exponential of the real part of the chiral loop function. 
In this way, this part of the loop function is resummed, while the imaginary one (the width) is kept unresummed in the denominator, $M_R^2-s-iM_R\Gamma_R(s)$. This violates 
analyticity at the next-to-next-to-leading order in the chiral expansion, which is a small effect numerically. The unitarized scalar form factor \cite{JOP}, which is a solution 
of the Muskelishivili-Omn\`es problem for three channel case ($K\pi$, $K\eta$, $K\eta^\prime$) will be employed.
 \item Dispersive representation (BEJ): A three-times subtracted dispersion relation will be used for the vector form factor. The input phaseshift will be defined using the 
vector form factor in eq.~(\ref{FFs with short distance constraints})
\begin{equation}\label{phaseshift}
\mathrm{tan}\,\delta(s)\,=\,\frac{Im\left[\tilde{f}_+^{K^-\eta}(s)\right]}{Re\left[\tilde{f}_+^{K^-\eta}(s)\right]},\,
\end{equation}
so that the output form factor is
\begin{equation}
 \widetilde{f}_+(s)\,=\,\mathrm{exp}\left[\lambda^\prime_+\frac{s}{m_\pi^2}+\frac{1}{2}\lambda^{\prime\prime}_+\frac{s^2}{m_\pi^4}+\frac{s^3}{\pi}\int_{s_{K\pi}}^{s_{cut}}ds^\prime
\frac{\delta(s^\prime)}{(s^\prime)^3(s^\prime-s-i0)}\right]\,,
\end{equation}
where $s_{K\pi}=(m_K+m_\pi)^2$. $\Gamma_{K^\star}(s)$ includes only the $K\pi$ cut and the whole complex loop function is resummed in the denominator, which keeps 
analyticity exactly. This vector form factor neglects inelastic coupled channel effects, which is in principle a questionable approximation. We anticipate that the agreement 
with data does not call for including these effects at the moment. The slope parameters, $\lambda^{\prime(\prime)}_+$, encode the very low energy dynamics. The unitarized 
scalar form factor will also be used \cite{JOP}.\\
\end{itemize}

\section{Predicting the $\tau^-\to K^-\eta\nu_\tau$ decays}\label{Pred Keta}
\vspace*{0.25cm}
Eqs.~(\ref{FFs with short distance constraints}) also hold for the $\widetilde{f}_{+,0}^{(K\pi)^-}(s)$ form factors. Therefore, the detailed knowledge of the $K\pi$ form 
factors~\cite{Kpi} could in principle be used to predict the the $\tau^-\to K^-\eta^{(\prime)}\nu_\tau$ decays. While this is true for the vector form factors, the unitarization 
procedure of the scalar form factors breaks this universality and different $\widetilde{f}_{0}^{K^-\,\pi^0/\eta/\eta^\prime}(s)$ are obtained. Then, the unitarization procedure 
can be tested through the effect of the scalar form factors in the $\tau^-\to K^-\eta^{(\prime)}\nu_\tau$ decays. While the suppression of the scalar contribution in the $\eta$ 
case makes it difficult to check finely the unitarized results, its leading role in the $\eta^\prime$ case would give us un opportunity to probe the $\widetilde{f}_{0}^{K^-\eta^\prime}(s)$ 
form factor.\\

Taking this discussion into account, we have predicted the $\tau^-\to K^-\eta\nu_\tau$ decays (and later on the $\tau^-\to K^-\eta^\prime\nu_\tau$ decays in section \ref{Ketap})
as explained below:
\begin{itemize}
 \item In the dipole model, we have taken the $K^\star(892)$, $K^\star(1410)$ and $K_0^\star(1430)$ mass and width from the PDG \cite{Beringer:1900zz} because Breit-Wigner 
parametrizations are employed in this reference. We estimated the relative weight of them using $\gamma=\frac{F_VG_V}{F^2}-1=-0.021\pm0.031$.
 \item In the JPP parametrization, we have used the best fit results of the second reference in \cite{Kpi} for the vector form factor. The scalar form factor was obtained 
from the solutions (6.10) and (6.11) of the second reference in \cite{JOP}. The scalar form factors have also been treated alike in the BEJ approach.
 \item In the BEJ representation, one would use the best fit results of the last reference in \cite{Kpi} to obtain our vector form factor. However, the slope form factor parameters, 
$\lambda_+^\prime$ and $\lambda_+^{\prime\prime}$, are very sensitive to isospin breaking corrections on the particle masses. For this reason we estimated the corresponding 
parameters for the $K^-\pi^0$ case of interest for the $\tau^-\to K^-\eta^{(\prime)}\nu_\tau$ decays. We have therefore used the results in the middle column of table \ref{Tab:Fake fit} 
to predict the $\tau^-\to K^-\eta^{(\prime)}\nu_\tau$ decays. More details on this procedure can be found in Ref.~\cite{OurWork}.\\

\begin{table}[h!]
\caption{\label{Tab:Fake fit}\small{Results for the fit to Belle $\tau^-\to K_S\pi^-\nu_\tau$ data \cite{BfactKpi} with a three-times subtracted dispersion relation including two vector 
resonances in $\widetilde{f}_+^{K\pi}(s)$, according to eq.~(\ref{FFs with short distance constraints}), and resumming the loop function in the denominator; as well as the 
unitarized scalar form factor \cite{JOP}. The middle column is obtained using the masses of the $K^-$ and $\pi^0$ mesons and the last column using the $K_S$ and $\pi^-$ 
masses actually corresponding to the data.}}
\begin{center}
\begin{tabular}{ccc}
\br
Parameter& Best fit with $K^-\pi^0$ masses & Best fit with $K_S\pi^-$ masses\\
\mr
$\lambda_+^\prime\times 10^{3}$&$22.2\pm0.9$& $24.7\pm0.8$\\
$\lambda_+^{\prime\prime}\times 10^{4}$&$10.3\pm0.2$& $12.0\pm0.2$\\
$M_{K^\star}$ (MeV)&$892.1\pm0.6$& $892.0\pm0.9$\\
$\Gamma_{K^\star}$ (MeV)&$46.2\pm0.5$& $46.2\pm0.4$\\
$M_{K^{\star\prime}}$ (GeV)&$1.28\pm0.07$& $1.28\pm0.07$\\
$\Gamma_{K^{\star\prime}}$ (GeV)&$0.16^{+0.10}_{-0.07}$& $0.20^{+0.06}_{-0.09}$\\
$\gamma$&$-0.03\pm0.02$& $-0.04\pm0.02$\\
\br
\end{tabular}
\end{center}
\end{table}

\end{itemize}
Using these inputs we have found the differential decay distributions for the three different approaches considered using eq.~(\ref{spectral function}). 
This one is, in turn, related to the experimental data by using 
\begin{equation}\label{theory_to_experiment}
 \frac{dN_{events}}{dE}\,=\,\frac{d\Gamma}{dE}\frac{N_{events}}{\Gamma_\tau BR(\tau^-\to K^-\eta\nu_\tau)}\Delta E_{bin}\,.
\end{equation}

\begin{figure}[h!]
\begin{center}
\vspace*{1.25cm}
\includegraphics[scale=0.75]{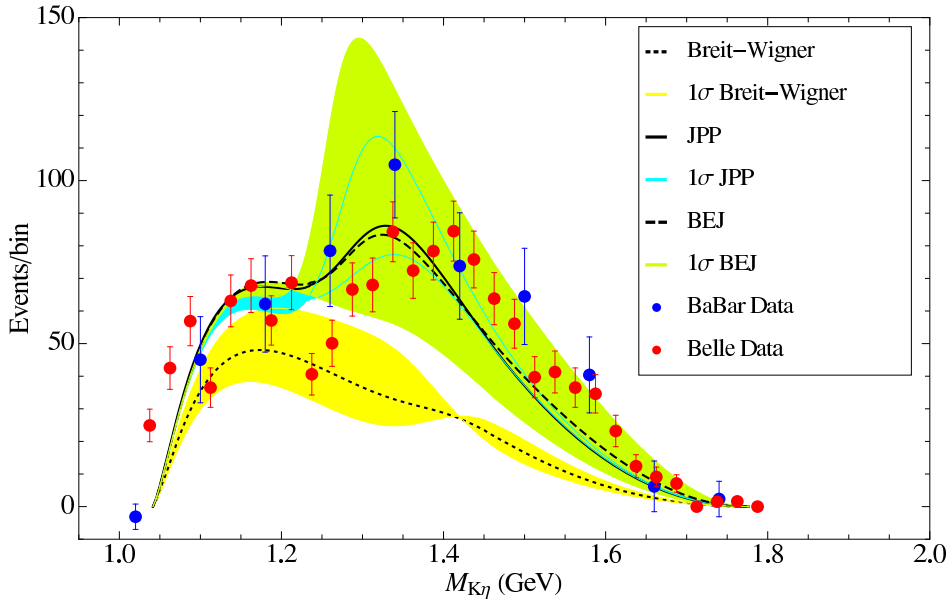}
\caption{\label{fig:Pred_Keta} \small{BaBar (blue) and Belle (red) \cite{BfactKeta} data for the $\tau^-\to K^-\eta\nu_\tau$ decays are confronted to the predictions obtained 
in the BW (dotted), JPP (solid) and BEJ (dashed) approaches (see the main text for details) which are shown together with the corresponding one-sigma error bands in yellow, 
light blue and light green, respectively.}}
\end{center}
\end{figure}

In Fig.\ref{fig:Pred_Keta} we show our predictions based on the $K\pi$ system according to BW, JPP and BEJ. In this figure we have normalized the BaBar data 
to Belle's using eq.~(\ref{theory_to_experiment}). We notice some tension between the BaBar and Belle data sets and strange oscillations of some Belle points that 
may hint to a systematic issue or an underestimation of the errors. In this figure, the one-sigma contours for the three approaches are also shown. The 
corresponding branching ratios are displayed in table \ref{Tab:Pred_Keta}, where the $\chi^2/dof$ is also given. It is seen that the BW approximation fails both in 
the decay width and the differential decay distribution shape. On the contrary, the JPP and BEJ treatments give already a good agreement with both of them. These 
results point to the BW modelization being a too rough approximation to the underlying dynamics. We understand this fact since, as we have discussed, this approach 
does not resum FSI and violates analyticity and unitarity severely. We will check this conclusion in the next section by fitting BaBar and Belle data to discard the 
possibility that the input parameters for the BW prediction were inappropriate.\\

\begin{table}[h!]
 \caption{\small{Predicted branching ratio of the $\tau^-\to K^-\eta\nu_\tau$ decays according to the different approaches used. The corresponding $\chi^2/dof$ values are 
also shown and the PDG branching fraction is given for reference.}}\label{Tab:Pred_Keta}
\begin{center}
\begin{tabular}{ccc}
\br
Source& Branching ratio & $\chi^2/dof$\\
\mr
Dipole Model (BW)&$\left(0.78^{+0.17}_{-0.10}\right)\cdot 10^{-4}$& $8.3$\\
JPP&$\left(1.47^{+0.14}_{-0.08}\right)\cdot 10^{-4}$& $1.9$\\
BEJ&$\left(1.49\pm0.05\right)\cdot 10^{-4}$& $1.5$\\
Experimental value&$\left(1.52\pm0.08\right)\cdot 10^{-4}$& -\\
\br
\end{tabular}
\end{center}
\end{table}

\section{Fitting the $\tau^-\to K^-\eta\nu_\tau$ data}
\vspace*{0.25cm}
We have considered different kinds of fits to BaBar and Belle $\tau^-\to K^-\eta\nu_\tau$ data. We have first assessed, in full generality, that these decays are insensitive 
to the mass and width of the $K^\star(892)$ resonance, as it could be expected since phase space for the $K\eta$ channel opens above the region of $K^\star(892)$ dominance. 
This is even more the case for the slope parameters of BEJ, which are determined by the Physics immediately above the $K\pi$ threshold. For this reason, we have borrowed 
these parameters from the corresponding predictions used in the previous section. We have thus fitted only the $K^\star(1410)$ parameters in all three approaches \footnote{The 
dependence on the $K^\star_0(1430)$ mass and width in the dipole (BW) model is very mild and can be neglected.}.\\

\begin{figure}[h!]
\begin{center}
\vspace*{1.25cm}
\includegraphics[scale=0.75]{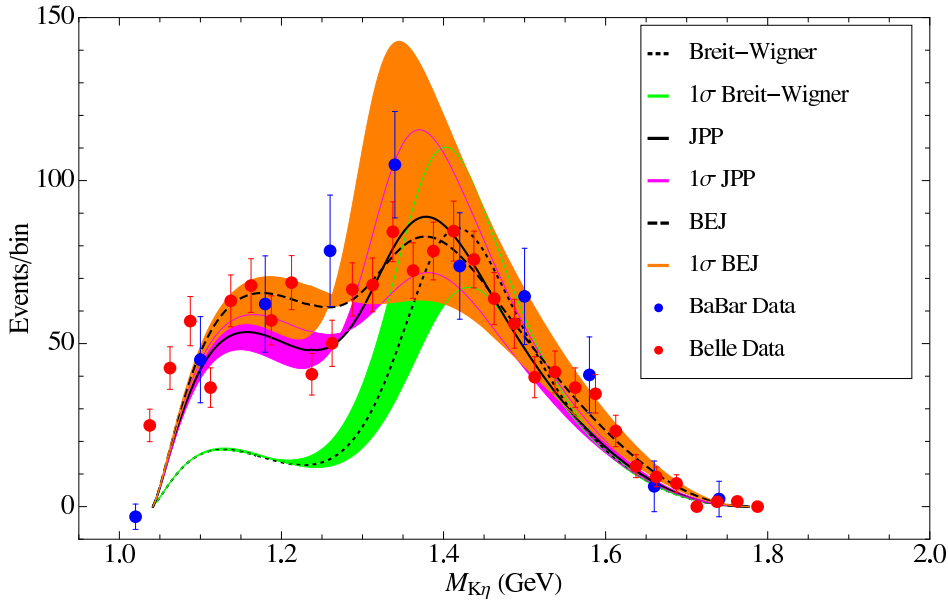}
\caption{\label{fig:Fit_Keta} \small{BaBar (blue) and Belle (red) \cite{BfactKeta} data for the $\tau^-\to K^-\eta\nu_\tau$ decays are confronted to the best fit results 
obtained in the BW (dotted), JPP (solid) and BEJ (dashed) approaches (see the main text for details) which are shown together with the corresponding one-sigma error bands in 
light green, pink and orange, respectively. The BW curve corresponds to the one-parameter fit while the JPP and BEJ ones correspond to three-parameter fits.}}
\end{center}
\end{figure}

Our best fit results for the branching ratios are given in table \ref{Tab:Fit_Keta}, including the corresponding $\chi^2/dof$. These results are obtained with the best fit 
parameter values shown in table \ref{Tab:Fit_results}, which can be compared to the reference values used for the predictions in the previous section (these are recalled in 
table \ref{Tab:Fit_Reference}). The corresponding decay distributions including one-sigma error bands are plotted in Fig.~\ref{fig:Fit_Keta}.\\

These results show that the BW model does not approximate the underlying physics for any value of its parameters and should be discarded. On the contrary, JPP and BEJ yield 
good fits to data with $\chi^2/dof$ values close to unity. This confirms that the simplified treatment of FSI in BW, which misses the real part of the two-meson 
rescatterings, violates analyticity by construction and does not resum FSI, is responsible for the failure.\\

\begin{table}[h!]
\caption{\label{Tab:Fit_Keta} \small{The branching ratios and $\chi^2/dof$ obtained in BW, JPP and BEJ fitting $\gamma$ only and also the $K^\star(1410)$ parameters are 
displayed. Other parameters were fixed to the reference values used in section \ref{Pred Keta}. The PDG branching fraction is also given for reference.}}
\begin{center}
\begin{tabular}{ccc}
\br
Source & Branching ratio & $\chi^2/dof$\\
\mr
Dipole Model (BW) (Fit $\gamma$)&$\left(0.96^{+0.21}_{-0.15}\right)\cdot10^{-4}$& $5.0$\\
Dipole Model (BW) (Fit $\gamma$, $M_{K^{\star\prime}}$, $\Gamma_{K^{\star\prime}})$ &Unphysical result& -\\
JPP (Fit $\gamma$)&$\left(1.50^{+0.19}_{-0.11}\right)\cdot 10^{-4}$& $1.6$\\
JPP (Fit $\gamma$, $M_{K^{\star\prime}}$, $\Gamma_{K^{\star\prime}})$&$\left(1.42\pm0.04\right)\cdot 10^{-4}$& $1.4$\\
BEJ (Fit $\gamma$)&$\left(1.59^{+0.22}_{-0.16}\right)\cdot 10^{-4}$& $1.2$\\
BEJ (Fit $\gamma$, $M_{K^{\star\prime}}$, $\Gamma_{K^{\star\prime}})$ &$\left(1.55\pm0.08\right)\cdot 10^{-4}$& $0.8$\\
Experimental value&$\left(1.52\pm0.08\right)\cdot 10^{-4}$& -\\
\br
\end{tabular}
\end{center}
\end{table}

\begin{table}[h!]
\caption{\label{Tab:Fit_results} \small{The best fit parameter values corresponding to the different alternatives considered in table \ref{Tab:Fit_Keta} are given. These can 
be compared to the reference values, which are given in table \ref{Tab:Fit_Reference}. BEJ results for the mass and width of the $K^{\star}(1410)$ correspond to pole values, 
while JPP figures are given for the model parameter as in the original literature.}}
\begin{center}
\begin{tabular}{cccccccc}
\br
\backslashbox{Fitted value}{Approach}&Dipole Model (BW)&JPP&BEJ\cr
\mr
$\gamma$&$-0.174\pm0.007$&$-0.063\pm0.007$&$-0.041\pm0.021$\cr
\mr
$\gamma$&Unphysical&$-0.078^{+0.012}_{-0.014}$&$-0.051^{+0.012}_{-0.036}$\cr
$M_{K^{\star'}}$ (MeV) &best fit&$1356\pm11$&$1327^{+30}_{-38}$\cr
$\Gamma_{K^{\star'}}$ (MeV) &parameters&$232^{+30}_{-28}$&$213^{+72}_{-118}$\cr
\br
\end{tabular}
\end{center}
\end{table}

\begin{table}[h!]
\caption{\label{Tab:Fit_Reference} \small{Reference values (used in section \ref{Pred Keta}) corresponding to the best fit parameters appearing in table \ref{Tab:Fit_results}. 
Again BEJ results are pole values and JPP ones are model parameters. The latter are converted to resonance pole values at the end of this section and the results are discussed 
in section \ref{Concl}.}}
\begin{center}
\begin{tabular}{cccccccc}
\hline
\backslashbox{Reference value}{Approach}&Dipole Model (BW)&JPP&BEJ\cr
\hline
$\gamma$&$-0.021\pm0.031$&$-0.043\pm0.010$&$-0.029\pm0.017$\cr
$M_{K^{\star'}}$ (MeV) & $1414\pm15$ &$1307\pm17$&$1283\pm65$\cr
$\Gamma_{K^{\star'}}$ (MeV) & $232\pm21$ &$206\pm49$&$163\pm68$\cr
\hline
\end{tabular}
\end{center}
\end{table}

These results are plotted in Fig.~\ref{fig:Fit_Keta}. Looking at the JPP and BEJ results in more detail one can notice that:
\begin{itemize} 
 \item The $\chi^2/dof$ of both approaches improves by $15\leftrightarrow 20\%$ fitting only $\gamma$. Fitted and reference values are consistent (see table \ref{Tab:Fit_Reference}).
Both the $\tau^-\to (K\pi)^-\nu_\tau$ and the $\tau^-\to K^-\eta\nu_\tau$ decays are sensitive to the interplay between the first two vector resonances and this agreement is a 
good autoconsistency test.
 \item Fitting also the $K^\star(1410)$ parameters improves the results by $\sim13\%$ in JPP and by $\sim33\%$ in BEJ. The three-parameter fits do not yield physical results 
in BW. Specifically, $K^\star(1410)$ mass and width approach to the $K^\star(892)$ values and $|\gamma|$ is one order of magnitude larger than the reference values, which makes 
us discard this result. Although the branching ratios of both JPP and BEJ agree with the PDG value, the JPP branching ratios are closer to its lower limit while BEJ is nearer 
the upper one. Deviations of the three-parameter best fit values with respect to the input values lie within errors in BEJ, as it so happens with $\Gamma_{K^{\star\prime}}$ 
in JPP. However, there are small tensions between the reference and best fit values of $M_{K^{\star\prime}}$ and $\gamma$ in JPP.\\
\end{itemize}

Although the BW curve in Fig.~\ref{fig:Fit_Keta} has improved with respect to Fig.~\ref{fig:Pred_Keta} and seems to agree well with the data in the 
higher-energy half of the spectrum, it fails completely at lower energies. On the contrary, JPP and BEJ provide good quality fits to data which are satisfactory along the 
whole phase space. Even though the vector form factor gives the dominant contribution to the decay width, the scalar form factor is not negligible and gives 
$\sim(3\leftrightarrow4)\%$ of the branching fraction in the JPP and BEJ cases. In the BW model this contribution is $\sim7\%$.\\

We have then translated the JPP model values appearing in tables \ref{Tab:Fit_results} and \ref{Tab:Fit_Reference} to the physical pole values \cite{Escribano:2002iv}, 
yielding $M_{K^{\star\prime}}=1332^{+16}_{-18}\,,\,\Gamma_{K^{\star\prime}}=220^{+26}_{-24}$ (best fit values) and 
$M_{K^{\star\prime}}=1286^{+26}_{-28}\,,\,\Gamma_{K^{\star\prime}}=197^{+41}_{-45}$ (reference values), with energy units given in MeV. Remarkable 
agreement is found between our best fit values in the JPP and BEJ cases, since the latter yields $M_{K^{\star\prime}}=1327^{+30}_{-38}\,,\,\Gamma_{K^{\star\prime}}=213^{+72}_{-118}$.\\

\section{Predicting the $\tau^-\to K^-\eta^\prime\nu_\tau$ decays}\label{Ketap}
\vspace*{0.25cm}
We can finally predict the $\tau^-\to K^-\eta^\prime\nu_\tau$ decays. In this case, the good understanding of the tau decays into $K\pi$ and $K\eta$ processes can only be exploited 
to a limited extent in the $K\eta^\prime$ decays. This is because while the vector form factor dominates the former decays, the scalar one essentially saturates the contribution to the 
latter. Therefore, this prediction will be more a test of the unitarization results obtained for the corresponding scalar form factor. The only information that one has for this 
decay channel is the upper limit fixed at ninety percent confidence level by the BaBar Collaboration \cite{Lees:2012ks}, $BR<4.2\cdot10^{-6}$. We will see that our predictions 
respect this bound and hint to the soon discovery of this decay channel at Belle-II.\\

\begin{figure}[h!]
\begin{center}
\vspace*{1.25cm}
\includegraphics[scale=0.75]{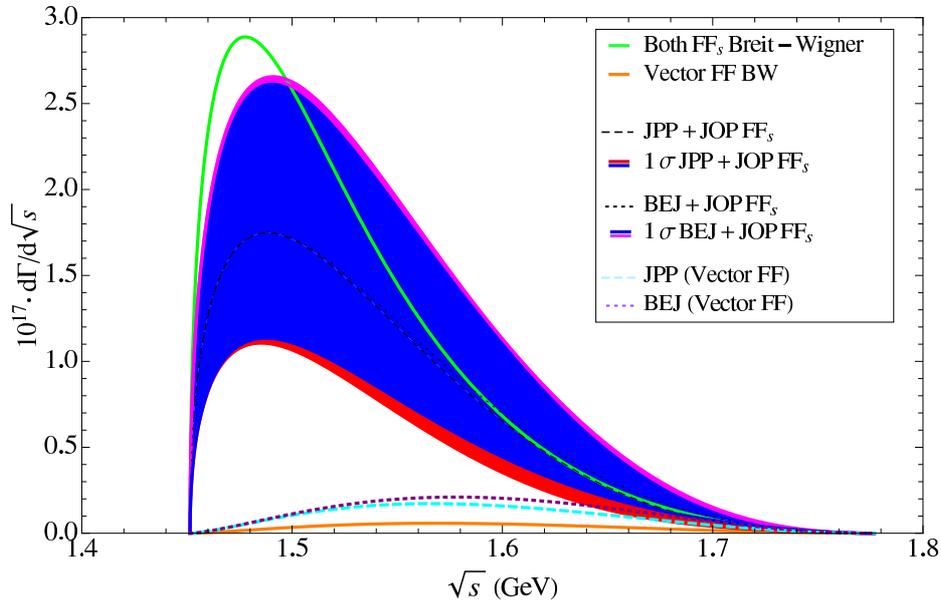}
\caption[]{\label{fig:Ketap} \small{The predicted $\tau^-\to K^-\eta^\prime\nu_\tau$ decay width according to BW (green, its big uncertainty is not shown for clarity of the 
figure), JPP (blue with lower band in red) and BEJ (blue with upper part in pink) is shown. In these last two the scalar form factor corresponds to Ref.~\cite{JOP}, 
which is represented by the author's initials, JOP, in the figure's legend. The corresponding vector form factor contributions, which are subleading are plotted 
in orange (solid), blue (dashed) and purple (dotted).}}
\end{center}
\end{figure}

We have ellaborated these predictions using our best fit results in the BW (one-parameter fit) JPP and BEJ (three-parameter fits) cases and the unitarized scalar form factors 
in the last two approaches. Our results are plotted in Fig.~\ref{fig:Ketap} and the branching ratios shown in table \ref{Tab:Pred_Ketap}. It is seen that the decay width is 
indeed dominated by the scalar form factor, with the vector one contributing in the interval $(9\leftrightarrow15)\%$ to the corresponding decay width. The BW prediction is 
only shown for reference, but its associated (large) error bands are not shown for clarity. Since the same scalar form factor is used JPP and BEJ and it basically saturates the 
decay width, the differences between them are tiny.\\

\begin{table*}[h!]
\caption{\label{Tab:Pred_Ketap} \small{Predicted branching ratios for the $\tau^-\to K^-\eta^\prime\nu_\tau$ decays. The BaBar upper limit is also shown \cite{Lees:2012ks}.}}
\begin{center}
\begin{tabular}{cc}
\hline 
Source & Branching ratio\\
\hline
Dipole Model (BW) (Fit)&$(1.45^{+3.80}_{-0.87})\cdot10^{-6}$\\
JPP (Fit)&$(1.00^{+0.37}_{-0.29})\cdot10^{-6}$\\
BEJ (Fit)&$(1.03^{+0.37}_{-0.29})\cdot10^{-6}$\\
Experimental bound&$<4.2\cdot10^{-6}$ at $90\%$ C.L.\\
\hline
\end{tabular}
\end{center}
\end{table*}

\section{Conclusions and outlook}\label{Concl}
\vspace*{0.25cm}
Hadronic tau decays constitute and ideal tool to study the hadronization of QCD currents in a clean environment. Here we have reported our study of the $\tau^-\to K^-\eta^{(\prime)}\nu_\tau$ 
decays which was triggered by the recent measurements and searches performed by the BaBar and Belle Collaborations \cite{BfactKeta, Lees:2012ks}. The $K\eta$ channel is 
dominated by the vector form factor contribution, which can be predicted accurately on the basis of previous studies of the $K\pi$ system. In fact, although the information 
on the $K^\star(892)$ resonance needs to be borrowed from the $K\pi$ studies, the $K\eta$ decays are extremely sensitive to the characteristics of the $K^\star(1410)$ resonance, 
whose properties can therefore be determined with precission in this analysis.\\

We have proposed a description of these processes encoding the hadronization properties in the tilded form factors, which yield more compact and symmetric expressions than 
used previously. After deriving the participating vector and scalar form factors within $R \chi T$ we have discussed the treatment of FSI, which turns out to be 
crucial in the considered decays. The popular Breit-Wigner (dipole) parametrizations do not resum FSI and break analyticity at leading order by neglecting the real part of the 
chiral loop functions. Omn\`es-like resummations resum only the real part of these functions through their exponentiation which violates analyticity (slightly at the numerical 
level). Finally, a dispersive form factor resums the whole loop function and respects analyticity by construction.\\

We have found that the Breit-Wigner model fails dramatically in accounting for the data, while the Omn\`es-like resummation and the dispersive representation do provide good 
fits to data. Although on theory grounds the simple Breit-Wigner models are always poor approximations its eventual agreement with data in decays with an easy dynamics still 
motivates its wide use in the analysis of hadronic data. These results for the $\tau^-\to K^-\eta\nu_\tau$ decays show neatly that also phenomenology suggests not to employ them.\\

Our results for the $K^\star(1410)$ pole mass and width (in MeV) and interaction strength are
\begin{equation}
 M_{K^{\star\prime}}\,=\,1327^{+30}_{-38},\quad\Gamma_{K^{\star\prime}}\,=\,213^{+72}_{-118},\quad \gamma\,=\,-0.051^{+0.012}_{-0.036}\,,
\end{equation}
in the dispersive representation (BEJ) and 
\begin{equation}
 M_{K^{\star\prime}}\,=\,1332^{+16}_{-18},\quad\Gamma_{K^{\star\prime}}\,=\,220^{+26}_{-24},\quad \gamma\,=\,-0.078^{+0.012}_{-0.014}\,,
\end{equation}
in the exponential parametrization (JPP). Our determination of these parameters happens to be competitive with its traditional extraction from $\tau^-\to (K\pi)^-\nu_\tau$ 
decays. In order to illustrate this, we average the JPP and BEJ determinations from the $K\pi$ \cite{Kpi} and $K\eta$ systems, respectively, to find
\begin{equation}
 M_{K^{\star\prime}}\,=\,1277^{+35}_{-41},\quad\Gamma_{K^{\star\prime}}\,=\,218^{+95}_{-66},\quad \gamma\,=\,-0.049^{+0.019}_{-0.016}\,,
\end{equation}
from $K\pi$ and 
\begin{equation}
 M_{K^{\star\prime}}\,=\,1330^{+27}_{-41},\quad\Gamma_{K^{\star\prime}}\,=\,217^{+68}_{-122},\quad \gamma\,=\,-0.065^{+0.025}_{-0.050}\,,
\end{equation}
from $K\eta$, which opens an alternative way of determining these parameters. New, more precise data on the $\tau^-\to (K\pi)^-\nu_\tau$ and 
$\tau^-\to K^-\eta\nu_\tau$ decays will make possible a more accurate determination of these parameters. We are currently working \cite{InProgress} on a combined study of 
Belle's data on the $\tau^-\to K_S\pi^-\nu_\tau$ and $\tau^-\to K^-\eta\nu_\tau$ decays with the purpose of obtaining more accurate isospin averaged values for the slope, 
$K^\star(892)$ and $K^\star(1410)$ parameters. When BaBar's data for the $\tau^-\to K^-\pi^0\nu_\tau$ and $\tau^-\to K^-\eta\nu_\tau$ data become available the effect of 
isospin breaking corrections on these quantities could be studied too.\\

We thus provide TAUOLA with QCD-motivated currents for these processes and improve the understanding of the strange spectral function thanks to our more detailed knowledge 
of the $\tau^-\to K^-\eta^{(\prime)}\nu_\tau$ decays.\\

\subsection*{Acknowledgments}
I thank Sergi Gonz\'alez-Sol\'{\i}s for useful comments on the manuscript. The author benefited from a grant of the XIV Mexican Workshop on Particles and Fields covering his 
accommodation expenses and registration fee. This work has been partially funded by Conacyt and DGAPA. The support of project PAPIIT IN106913 is also acknowledged. The 
research reported here was supported in part by the FPI scholarship BES-2012-055371 (S.~G.-S.), the Ministerio de Ciencia e Innovaci\'on under grants FPA2011-25948 and 
AIC-D-2011-0818, the European Commission under the 7thFramework Programme through the “Research Infrastructures” action of the “Capacities” Programme Call: 
FP7-INFRA-STRUCTURES-2008-1 (Grant Agreement N. 227431), the Spanish Consolider-Ingenio 2010 Programme CPAN (CSD2007-00042), and the Generalitat de Catalunya under grant 
SGR2009-00894 (Rafel Escribano, S.~G.-S. and P.~R.).\\


\section*{References}
\begin{thebibliography}{99}
 \bibitem{HadTauDec}
  E.~Braaten, S.~Narison and A.~Pich,
  Nucl.\ Phys.\ B {\bf 373} (1992) 581.
  E.~Braaten,
  Phys.\ Rev.\ Lett.\  {\bf 60} (1988) 1606,
  Phys.\ Rev.\ D {\bf 39} (1989) 1458.
  E.~Braaten and C.~-S.~Li,
  Phys.\ Rev.\ D {\bf 42} (1990) 3888.
  S.~Narison and A.~Pich,
  Phys.\ Lett.\ B {\bf 211} (1988) 183.
  A.~Pich,
  Conf.\ Proc.\ C {\bf 890523} (1989) 416.
  M.~Davier, A.~Hocker and Z.~Zhang,
  Rev.\ Mod.\ Phys.\  {\bf 78} (2006) 1043.
A. Pich, 
``Precision Tau Physics'',
arXiv:1310.7922 [hep-ph], 
to be published in ``Progress in Particle and Nuclear Physics''.

\bibitem{InclTauDec}
  M.~Davier, S.~Descotes-Genon, A.~Hocker, B.~Malaescu and Z.~Zhang,
  Eur.\ Phys.\ J.\ C {\bf 56} (2008) 305.
  M.~Beneke and M.~Jamin,
  JHEP {\bf 0809} (2008) 044.
  A.~Pich,
 arXiv:1107.1123 [hep-ph]. Published in the Proc. of the Workshop on Precision Measurements of $\alpha_S$
 9-11 Feb 2011. Munich, Germany.
  D.~Boito, M.~Golterman, M.~Jamin, A.~Mahdavi, K.~Maltman, J.~Osborne and S.~Peris,
  Phys.\ Rev.\ D {\bf 85} (2012) 093015.
  R.~Barate {\it et al.}  [ALEPH Collaboration],
  Eur.\ Phys.\ J.\ C {\bf 11} (1999) 599.
  G.~Abbiendi {\it et al.}  [OPAL Collaboration],
  Eur.\ Phys.\ J.\ C {\bf 35} (2004) 437.
  K.~Maltman, C.~E.~Wolfe, S.~Banerjee, J.~M.~Roney and I.~Nugent,
  Int.\ J.\ Mod.\ Phys.\ A {\bf 23} (2008) 3191.
  M.~Antonelli, V.~Cirigliano, A.~Lusiani and E.~Passemar,
  arXiv:1304.8134 [hep-ph].
  K.~G.~Chetyrkin, J.~H.~Kuhn and A.~A.~Pivovarov,
  Nucl.\ Phys.\ B {\bf 533} (1998) 473
  A.~Pich and J.~Prades,
  JHEP {\bf 9910} (1999) 004.
  J.~G.~Korner, F.~Krajewski and A.~A.~Pivovarov,
  Eur.\ Phys.\ J.\ C {\bf 20} (2001) 259.
  J.~Kambor and K.~Maltman,
  Phys.\ Rev.\ D {\bf 62} (2000) 093023.
  S.~Chen, M.~Davier, E.~Gamiz, A.~Hocker, A.~Pich and J.~Prades,
  Eur.\ Phys.\ J.\ C {\bf 22} (2001) 31.
  E.~G\'amiz, M.~Jamin, A.~Pich, J.~Prades and F.~Schwab,
  JHEP {\bf 0301} (2003) 060.
  E.~G\'amiz, M.~Jamin, A.~Pich, J.~Prades and F.~Schwab,
  Phys.\ Rev.\ Lett.\  {\bf 94} (2005) 011803.
  P.~A.~Baikov, K.~G.~Chetyrkin and J.~H.~Kuhn,
  Phys.\ Rev.\ Lett.\  {\bf 95} (2005) 012003.
  E.~G\'amiz, M.~Jamin, A.~Pich, J.~Prades and F.~Schwab,
  PoS KAON {\bf } (2008) 008.

\bibitem{BfactKpi}
  B.~Aubert {\it et al.}  [BaBar Collaboration],
  Phys.\ Rev.\ D {\bf 76} (2007) 051104.
  D.~Epifanov {\it et al.}  [Belle Collaboration],
  Phys.\ Lett.\ B {\bf 654} (2007) 65.

\bibitem{Kpi}
  M.~Jamin, A.~Pich and J.~Portol\'es,
  Phys.\ Lett.\ B {\bf 640} (2006) 176,
{\bf 664} (2008) 78.
  B.~Moussallam,
  Eur.\ Phys.\ J.\ C {\bf 53} (2008) 401.
  D.~R.~Boito, R.~Escribano and M.~Jamin,
  Eur.\ Phys.\ J.\ C {\bf 59} (2009) 821,
  JHEP {\bf 1009} (2010) 031.

\bibitem{OurWork}
  R.~Escribano, S.~Gonz\'alez-Sol\'{\i}s and P.~Roig,
  JHEP {\bf 1310} (2013) 039.

\bibitem{BfactKeta}
  P.~del Amo Sanchez {\it et al.}  [BaBar Collaboration],
  Phys.\ Rev.\ D {\bf 83} (2011) 032002.
  K.~Inami {\it et al.}  [Belle Collaboration],
  Phys.\ Lett.\ B {\bf 672} (2009) 209.

\bibitem{TAUOLA}
  S.~Jadach, J.~H.~K\"uhn and Z.~Was,
  Comput.\ Phys.\ Commun.\  {\bf 64} (1990) 275.
  S.~Jadach, Z.~Was, R.~Decker and J.~H.~K\"uhn,
  Comput.\ Phys.\ Commun.\  {\bf 76} (1993) 361.
  N.~Davidson, G.~Nanava, T.~Przedzinski, E.~Richter-Was and Z.~Was,
  Comput.\ Phys.\ Commun.\  {\bf 183} (2012) 821.

\bibitem{Olga}
  O.~Shekhovtsova, T.~Przedzinski, P.~Roig and Z.~Was,
  Phys.\ Rev.\ D {\bf 86} (2012) 113008.
I.~M.~Nugent, T.~Przedzinski, P.~Roig, O.~Shekhovtsova and Z.~Was,
  Phys.\ Rev.\ D {\bf 88} (2013) 093012.

\bibitem{TauEta}
  A.~Pich,
  Phys.\ Lett.\ B {\bf 196} (1987) 561.
  E.~Braaten, R.~J.~Oakes and S.~-M.~Tse,
  Int.\ J.\ Mod.\ Phys.\ A {\bf 5} (1990) 2737.
  B.~A.~Li,
  Phys.\ Rev.\ D {\bf 55} (1997) 1436.
  G.~J.~Aubrecht, II, N.~Chahrouri and K.~Slanec,
  Phys.\ Rev.\ D {\bf 24} (1981) 1318.

\bibitem{Actis:2010gg}
 S.~Actis {\it et al.},
 Eur.\ Phys.\ J.\  C {\bf 66} (2010) 585.

\bibitem{Lees:2012ks}
  J.~P.~Lees {\it et al.}  [BaBar Collaboration],
  Phys.\ Rev.\ D {\bf 86} (2012) 092010.

\bibitem{Erler:2002mv}
  J.~Erler,
  Rev.\ Mex.\ Fis.\  {\bf 50} (2004) 200.

\bibitem{Ambrosino:2006gk}
  F.~Ambrosino {\it et al.}  [KLOE Collaboration],
  Phys.\ Lett.\ B {\bf 648} (2007) 267.

\bibitem{Antonelli:2010yf}
  M.~Antonelli, V.~Cirigliano, G.~Isidori, F.~Mescia, 
 {\it et al.},
  Eur.\ Phys.\ J.\ C {\bf 69} (2010) 399.

\bibitem{ChPT}
  S.~Weinberg,
  Physica A {\bf 96} (1979) 327.
  J.~Gasser and H.~Leutwyler,
  Annals Phys.\  {\bf 158} (1984) 142.
  J.~Gasser and H.~Leutwyler,
  Nucl.\ Phys.\ B {\bf 250} (1985) 465.

\bibitem{ChPT2}
  J.~Bijnens, G.~Colangelo and G.~Ecker,
  JHEP {\bf 9902} (1999) 020,
  Annals Phys.\  {\bf 280} (2000) 100.
  J.~Bijnens, L.~Girlanda and P.~Talavera,
  Eur.\ Phys.\ J.\ C {\bf 23} (2002) 539.

\bibitem{Colangelo:1996hs}
G.~Colangelo, M.~Finkemeier and R.~Urech,
Phys.\ Rev.\ D {\bf 54} (1996) 4403.

\bibitem{LargeNc}
  G.~'t Hooft,
  Nucl.\ Phys.\ B {\bf 72} (1974) 461,
 {\bf 75} (1974) 461.
  E.~Witten,
  Nucl.\ Phys.\ B {\bf 160} (1979) 57.

\bibitem{LargeNc2}
A.~V.~Manohar,
 Published in 'Les Houches 1997, Probing the standard model of particle interactions, Pt. 2' 1091-1169.
 A.~Pich,
 Published in 'Tempe 2002, Phenomenology of large $N_C$ $QCD$' 239-258.
  V.~Cirigliano, G.~Ecker, H.~Neufeld and A.~Pich,
  JHEP {\bf 0306} (2003) 012.

\bibitem{RChT}
  G.~Ecker, J.~Gasser, A.~Pich and E.~de Rafael,
  Nucl.\ Phys.\ B {\bf 321} (1989) 311.
  G.~Ecker, J.~Gasser, H.~Leutwyler, A.~Pich and E.~de Rafael,
  Phys.\ Lett.\ B {\bf 223} (1989) 425.
 V.~Cirigliano, G.~Ecker, M.~Eidem\"uller, R.~Kaiser, A.~Pich and J.~Portol\'es,
 Nucl.\ Phys.\  B {\bf 753} (2006) 139.
  K.~Kampf and J.~Novotny,
  Phys.\ Rev.\ D {\bf 84} (2011) 014036.

\bibitem{GFs}
 P.~D.~Ruiz-Femen\'{\i}a, A.~Pich and J.~Portol\'es,
 JHEP {\bf 0307} (2003) 003.
 V.~Cirigliano, G.~Ecker, M.~Eidem\"uller, A.~Pich and J.~Portol\'es,
 Phys.\ Lett.\ B {\bf 596} (2004) 96.
 V.~Cirigliano, G.~Ecker, M.~Eidem\"uller, R.~Kaiser, A.~Pich and J.~Portol\'es,
 JHEP {\bf 0504} (2005) 006.
  P.~Roig and J.~J.~Sanz-Cillero,
  arXiv:1312.6206 [hep-ph].

\bibitem{HadTauDec2}
  F.~Guerrero and A.~Pich,
  Phys.\ Lett.\  B {\bf 412} (1997) 382.
  A.~Pich, J.~Portol\'es,
  Phys.\ Rev.\  {\bf D63 } (2001)  093005.
  J.~J.~Sanz-Cillero and A.~Pich,
  Eur.\ Phys.\ J.\ C {\bf 27} (2003) 587.
  D.~Gomez Dumm, A.~Pich and J.~Portol\'es,
  Phys.\ Rev.\ D {\bf 69} (2004) 073002.
  Z.~-H.~Guo,
  Phys.\ Rev.\ D {\bf 78} (2008) 033004.
  D.~G.~Dumm, P.~Roig, A.~Pich and J.~Portol\'es,
  Phys.\ Rev.\ D {\bf 81} (2010) 034031,
  Phys.\ Lett.\ B {\bf 685} (2010) 158.
  Z.~-H.~Guo and P.~Roig,
  Phys.\ Rev.\ D {\bf 82} (2010) 113016.
  D.~G.~Dumm and P.~Roig,
  Phys.\ Rev.\ D {\bf 86} (2012) 076009,
  Eur.\ Phys.\ J.\ C {\bf 73} (2013) 2528.
  P.~Roig
,  arXiv:1301.7626 [hep-ph], Ph. D. Thesis, Universitat de Val\`encia, Servei de Publicacions, November 2010.
  P.~Roig, A.~Guevara and G.~L\'opez~Castro,
  Phys.\ Rev.\ D {\bf 88} (2013) 033007, 
  arXiv:1401.4099 [hep-ph].
  P.~Roig,
  arXiv:1307.8277 [hep-ph].
  A.~Celis, V.~Cirigliano and E.~Passemar,
  arXiv:1309.3564 [hep-ph].
  Y.~-H.~Chen, Z.~-H.~Guo and H.~-Q.~Zheng,
  arXiv:1311.3366 [hep-ph].

\bibitem{FFs}
 V.~Mateu and J.~Portol\'es,
 Eur.\ Phys.\ J.\  C {\bf 52} (2007) 325.
  R.~Escribano, P.~Masjuan and J.~J.~Sanz-Cillero,
  JHEP {\bf 1105} (2011) 094.
  Y.~-H.~Chen, Z.~-H.~Guo and H.~-Q.~Zheng,
  Phys.\ Rev.\ D {\bf 85} (2012) 054018.
  L.~Y.~Dai, J.~Portol\'es and O.~Shekhovtsova,
  Phys.\ Rev.\ D {\bf 88} (2013) 056001.

\bibitem{JOP}
  M.~Jamin, J.~A.~Oller and A.~Pich,
  Nucl.\ Phys.\ B {\bf 587} (2000) 331, 
{\bf 622} (2002) 279,
  Eur.\ Phys.\ J.\ C {\bf 24} (2002) 237,
  Phys.\ Rev.\ D {\bf 74} (2006) 074009.

\bibitem{kappa}
  S.~Descotes-Genon and B.~Moussallam,
  Eur.\ Phys.\ J.\ C {\bf 48} (2006) 553.
  J.~Nebreda and J.~R.~Pel\'aez,
  Phys.\ Rev.\ D {\bf 81} (2010) 054035.
  S.~Prelovsek, T.~Draper, C.~B.~Lang, M.~Limmer, K.~-F.~Liu, N.~Mathur and D.~Mohler,
  Phys.\ Rev.\ D {\bf 82} (2010) 094507.
  Z.~-H.~Guo and J.~A.~Oller,
  Phys.\ Rev.\ D {\bf 84} (2011) 034005.
  M.~Doring and U.~G.~Meissner,
  JHEP {\bf 1201} (2012) 009.

\bibitem{BL}
  G.~P.~Lepage and S.~J.~Brodsky,
  Phys.\ Lett.\ B {\bf 87} (1979) 359,
  Phys.\ Rev.\ D {\bf 22} (1980) 2157.

\bibitem{GomezDumm:2000fz}
  D.~G\'omez Dumm, A.~Pich and J.~Portol\'es,
  Phys.\ Rev.\  D {\bf 62} (2000) 054014.

\bibitem{Beringer:1900zz}
  J.~Beringer {\it et al.}  [Particle Data Group Collaboration],
  Phys.\ Rev.\ D {\bf 86} (2012) 010001.

\bibitem{InProgress}
 R.~Escribano, S.~Gonz\'alez-Sol\'{\i}s, M.~Jamin and P.~Roig,
 work in progress.

\bibitem{Escribano:2002iv}
  R.~Escribano, A.~Gallegos, J.~L.~Lucio M, G.~Moreno and J.~Pestieau,
  Eur.\ Phys.\ J.\ C {\bf 28} (2003) 107.
\end{thebibliography}
\end{document}